%% file: main.tex
\documentclass[conference]{IEEEtran}
\IEEEoverridecommandlockouts
\usepackage{cite}
\usepackage{amsmath,amssymb,amsfonts,times,graphicx,booktabs,tabulary,multirow,overpic,xcolor,url}
\usepackage{algorithmic}
\usepackage{graphicx}
\usepackage{textcomp}
\usepackage{xcolor}
\usepackage[tableposition=top]{caption}
\usepackage{subcaption}
\def\BibTeX{{\rm B\kern-.05em{\sc i\kern-.025em b}\kern-.08em
    T\kern-.1667em\lower.7ex\hbox{E}\kern-.125emX}}
\begin{document}

\title{Dual-CLVSA: a Novel Deep Learning Approach to Predict Financial Markets with Sentiment Measurements}

\author{\IEEEauthorblockN{Jia Wang\IEEEauthorrefmark{1}, Hongwei Zhu\IEEEauthorrefmark{2}, Jiancheng Shen\IEEEauthorrefmark{3}, Yu Cao\IEEEauthorrefmark{1}, Benyuan Liu\IEEEauthorrefmark{1}}
\IEEEauthorblockA{\IEEEauthorrefmark{1}Department of Computer Science\\
\IEEEauthorrefmark{2}Department of Operations and Information Systems\\
University of Massachusetts Lowell\\
Email: \{jwang, ycao, bliu\}@cs.uml.edu, harry\_zhu@uml.edu}
\IEEEauthorblockA{\IEEEauthorrefmark{3}Dongwu Business School\\
Soochow University\\
Email: jcshen@suda.edu.cn}}

\maketitle

\begin{abstract}
It is a challenging task to predict financial markets. The complexity of this task is mainly due to the interaction between financial markets and market participants, who are not able to keep rational all the time, and often affected by emotions such as fear and ecstasy. Based on the state-of-the-art approach particularly for financial market predictions, a hybrid convolutional LSTM Based variational sequence-to-sequence model with attention (CLVSA), we propose a novel deep learning approach, named dual-CLVSA, to predict financial market movement with both trading data and the corresponding social sentiment measurements, each through a separate sequence-to-sequence channel. We evaluate the performance of our approach with backtesting on historical trading data of SPDR SP 500 Trust ETF over eight years. The experiment results show that dual-CLVSA can effectively fuse the two types of data, and verify that sentiment measurements are not only informative for financial market predictions, but they also contain extra profitable features to boost the performance of our predicting system.
\end{abstract}


\input{introduction.tex}

\input{related_work.tex}
\input{model_design.tex}
\input{exp_setting.tex}
\input{case_analysis}
\input{conclusion_futurework.tex}

\section*{Acknowledgment}
The authors wish to thank Richard Peterson, Managing Director at MarketPsych, for providing the authors with the proprietary Thomson Reuters MarketPsych Indices (TRMI) data.


\end{document}

%% file: introduction.tex
\section{Introduction}\label{Introduction}

Predicting financial markets is always challenging. The main difference between financial markets and other natural sequential events (e.g. DNA Sequences) is that, the evolution of financial markets is caused by the collective behavior of market participants rather than being governed by law of nature. The adaptive nature of financial markets makes their movement more complicated and difficult to predict as market participants are not able to be rational all the time. Once market participants are dominated by their emotions, such as fear, upset, ecstasy, and frustration, they inevitably cannot help overreacting or even making wrong decisions. Behavioral economists demonstrate that inefficiency of financial markets results from the spread of emotional responses among market participants, systematically biasing trading behaviors. As the group of market participants with the same emotion expands, their biased behaviors create trends of financial markets, which subsequently force the market price to move away from the true value. 

How to capture effective latent features from trading data is the key to build robust predicting systems for financial markets. Some research, such as \cite{kim2003financial,fernandez2003technical,dixon2016classification}, use machine learning algorithms (e.g. SVM, Nearest Neighborhood, and Feed-forward networks) to extract latent features from technical indicators. While technical indicators have been widely used by market participants, these methods may inevitably introduce human biases into models. Other popular sources for extracting latent features include market-related texts and information, such as reports, news, and tweets. Although classic economic theories believe that prices reflect all information, the sentiment data is still informative for traders due to a basic fact that people have emotions, and they take actions in markets. Moreover, many studies, such as \cite{yerkes1908relation,hebb1955drives, schwabe2009stress} have demonstrated that a person’s arousal level impacts decision-making . 

Therefore, if sentiment data can be obtained quickly, we will probably attain signals of the upcoming trend of financial markets. In this paper, we use Thomson Reuters MarketPsych Indices (TRMI) \cite{peterson2016trading} to investigate whether sentiment data provide signals that are more directional than random price movements. TRMI utilizes two groups of data sources to measure sentiment, namely, news and social media. The feed data consist of three types: a social media feed, a news media feed, and an aggregated feed of combined social and news media content. We proceed our research with the following three steps: 1. Verify the informativeness of TRMI data. We choose recurrent neural network with LSTM units as the baseline model ($LSTM_{s}$), and compare the expeirmental results on the three following datasets to examine the informativeness of TRMI data: historical trading data only, historical trading data with technical indicators, historical trading data with TRMI data. 2. Building upon the state-of-the-art experimental results of CLVSA on futures market predictions \cite{ijcai2019-514}, we evaluate this approach on historical SPDR SP500 Trust ETF (SPY)  trading data. Our experimental results show that CLVSA still achieves the best performance for historical SPY trading data, compared to baseline methods, such as $LSTM_{s}$. We thus use it as the baseline method of the 3rd step. 3. Design an effective method to fuse historical trading data and TRMI data based on the approach that is verified by the previous step. The intrinsic characteristics of historical trading data and TRMI data are so different that it does not work to directly fuse them at the input, which is verified by the first- and second-step experiment with SPY historical trading data with technical indicators and TRMI data. We design a fusion strategy, called dual-CLVSA, which applies two parallel channels of sequence-to-sequence framework for TRMI data and historical trading data to capture their distinctive features, and then combine the features to take advantage of the two different sources of information.

We summarize our contributions as follows:
\begin{enumerate}
\item Although there is complicated and implicit relevance between TRMI data and financial trading data, the different nature between the two types of data disturb them to fuse together with a simple manner. This paper explores fusion approaches to train TRMI data and raw financial trading data together.
\item We train our model using 8-year trading data of SPY with the corresponding TRMI data. Our experimental results show that our fusion model, dual-CLVSA, achieves the best performance on both financial and machine learning criteria, which also verifies that TRMI data contains extra informative features which can boost the performance of prediction systems.
\end{enumerate}

The remainder of the paper is organized as follows. Related work on financial market prediction with deep learning methods is presented in Section \ref{related work}. The methodology of our exploration of predicting financial markets with sentiment measurements is presented in Section \ref{model}. The data preprosessing and experimental setup and results are described in Section \ref{Experimental setup}. Two case studies are presented in Section \ref{Case Analysis}, followed by concluding remarks in Section \ref{Conclusion}.

%% file: related_work.tex
\section{Related Work}\label{related work}

Although traditional predicting approaches such as technical analysis/indicators have existed for over hundreds of years, automated trading systems based on pattern recognition and machine learning have become popular since the 1990s. Various algorithms, such as SVM, nearest-neighbour, decision trees, and feed-forward neural networks, have been applied to predict stocks, foreign exchange, and commodity futures markets \cite{kim2003financial,fernandez2003technical,rechenthin2014machine,dixon2016classification}. All the aforementioned work use technical indicators as input features. Since 2010s, more research start to utilize the power of deep learning algorithms to predict financial markets. \cite{ding2015deep,wang2018financial} use deep convolutional neural networks to capture potential trading features from financial events and financial trading data, respectively.
\cite{zhang2017stock} proposes a variant of LSTM enhanced by discrete fourier transform to discover Multi-Frequency Trading Patterns. \cite{bacoyannis2018idiosyncrasies} proposes an approach based on reinforcement learning to model automated data-centric decision makers in quantitative finance. 

Binding the local feature extraction ability of deep convolutional neural networks with the temporal features retention of LSTM, convolutional LSTM proposed by \cite{xingjian2015convolutional} has been applied in many fields such as weather forecasting \cite{xingjian2015convolutional}, image compression \cite{toderici2015variable}, and general algorithmic tasks (e.g. binary addition) \cite{kaiser2015neural}. The sequence-to-sequence framework proposed by \cite{sutskever2014sequence} achieves a significantly success in neural machine translation tasks, enhanced subsequently by inter-attention \cite{bahdanau2014neural} and self-attention \cite{cheng2016long}. \cite{kingma2013auto,rezende2014stochastic} propose variational auto-encoder (VAE) that uses the encoder to form the approximate posterior, then trains the generative decoder to approximate the inputs of the encoder with variational lower bound and KLD. SRM \cite{bayer2014learning,goyal2017z} extends the basic idea of VAE into recurrent networks, using backward recurrent neural networks as the approximate posterior instead. 

Some apporaches, such as \cite{ding2015deep,sun2017predicting,xu2018stock}, use natural language processing approaches to extract latent features within market-related texts and information, such as reports, news, and tweets. However, to the best of our knowledge, our research is among the first attempts to extract latent feature within sentiment measurements (e.g., Thomson Reuters MarketPsych Indices, a.k.a TRMI) with deep learning approaches. TRMI use natural language processing approaches to process sentiment-laden content in text, scoring content that pertains to specific companies, currencies, commodities, and countries. As the background of TRMI, varying levels of stress have been shown to map to cognitive performance in an inverse-U curve called the Yerkes-Dodson Law \cite{yerkes1908relation,hebb1955drives}. When stress levels are very high, complex problem-solving performance drops and reliance on pre-existing habits increases \cite{schwabe2009stress}. On the other hand, low stress levels also lead to subpar performance in complex decision-making environments due to inattention and slow reaction. Thus decision-makers typically perform with optimal cognition when arousal is in the middle of its range.

\begin{figure*}
\centering
\includegraphics[width=0.9\linewidth]{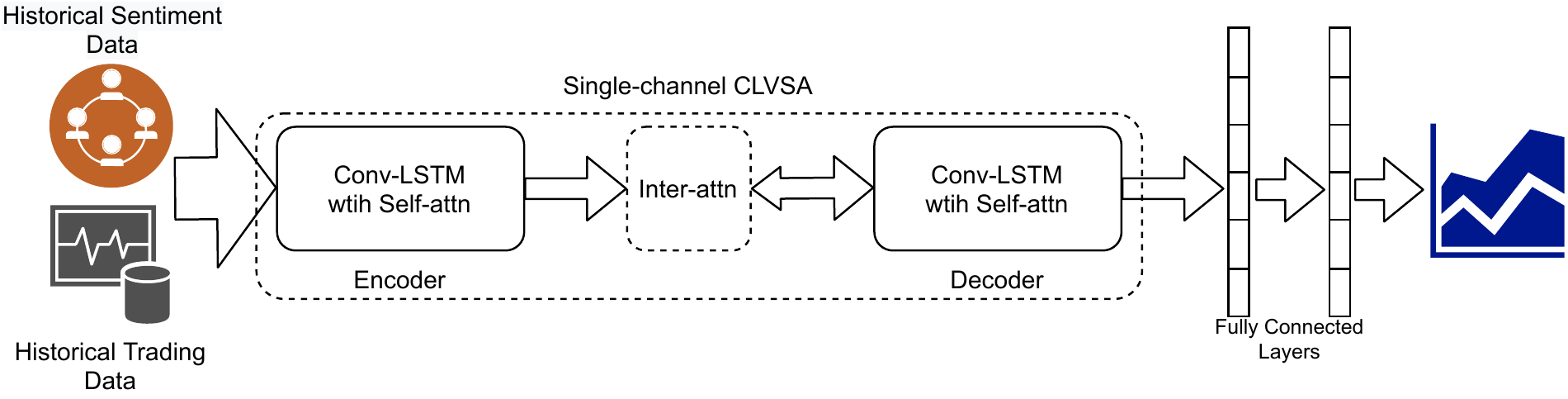}
\caption{The architecture of $CLVSA^2$. In this approach, we fuse historical sentiment data and trading data at the input, and our experimental results show that this fusion method does not work.}
\label{fig:spec_clvsa}
\end{figure*}

%% file: model_design.tex
\section{Methodology}\label{model}

\subsection{Introduction to Thomson Reuters MarketPsych Indices}
Thomson Reuters MarketPsych Indices (TRMI) measure the sentiment of market participants by distilling a massive collection of news and social media content through an extensive natural language processing framework. The indices consider different emotions (optimism, confusion, urgency etc.), as well as financial terms (interest rate, mergers etc.).

 TRMI have two groups of data sources: news and social media. The feed data consist of three types: a social media feed, a news media feed, and an aggregated feed of combined social and news media content. TRMI use natural language processing approaches to process sentiment-laden content in text, scoring content that pertains to specific companies, currencies, commodities, and countries. The entire content set includes over 2 million articles and posts daily from premium news wires, internet news sources, and social media. In our research, we focus on two types of TRMI: companies and equity index TRMI indices, and energy and material commodity TRMI indices. Each TRMI index consists of a combination of variables (PsychVars), such as AccountingBad, AccountingGood, Ambiguity, and Anger. Formally:
\begin{gather*}
BUZZ(a) = \sum_{c\in C(a), p \in P}|PsychVar_{c,p}|,
\end{gather*}
Where $Buzz(a)$ denotes the sum of the absolute values of all TRMI-contributing PsychVars. $P$ denotes the set of all PsychVars underlying any TRMI of the asset class, $C(a)$ denotes the set of all elements of asset $a$. For example, if $a$ is SP500, then  $C(a)$ represents the stocks of the 500 large companies in SP500. Each TRMI is then computed as a ratio of the sum of all relevant PsychVars to the Buzz. We define a function to indicate whether a PsychVar $p \in P$ is additive, subtractive, or irrelevant to a TRMI. Formally,
\begin{gather*}
I(t,p) =
\begin{cases}
+1, \text{ if } additive, \\
-1, \text{ if } subtractive, \\
0, \text{ if } irrelevant,
\end{cases}
\end{gather*}
\begin{gather*}
TRMI_{t}(a) = \frac{\sum_{c \in C(a), p \in P(t)}(I(t,p)*PsychVar_{c,p})}{Buzz(a)},
\end{gather*}
where $TRMI_{t}(a)$ denotes the $t-th$ TRMI for asset $a$. 

\subsection{Experimental Plan}
The main goal of our research is to verify our hypothesis that sentiment data can provide extra informative features to financial markets predictions. We thus design a three-step experimental plan based on the state-of-the-art model, CLVSA, with modifications as minimum as possible.
\begin{enumerate}
\item Verify the informativeness of TRMI data. We choose recurrent neural network with Long Short Term Memory (LSTM) units as the baseline model ($LSTM_{s}$), and use four different datasets to train $LSTM_{s}$, including SPDR SP500 Trust ETF (SPY) historical trading data only, SPY historical trading data with technical indicators, SPY historical trading data with the corresponding TRMI data, and SPY historical trading data with technical indicators and the corresponding TRMI data. We follow the methods in \cite{ijcai2019-514} to generate technical indicators. 
\item Identify a high performance baseline model with historical SPY trading data. In the previous research, our experimental results verify that CLVSA outperforms $LSTM_{s}$. If we could reproduce similar results for SPY historical trading data, CLVSA would qualify as the baseline model for the experiments with TRMI data. 
\item Explore effective methods to fuse historical trading data and TRMI data for financial market prediction. The intrinsic mechanisms of historical trading data and TRMI data are so different that it does not work to directly fuse them. We design a novel fusion method, named dual-CLVSA to address this problem.
\end{enumerate}

\subsection{dual-CLVSA: the fusion method}
The base approach of our fusion method, CLVSA, is a hybrid model consisting of convolutional LSTM units, sequence-to-sequence framework with attention mechanism, and stochastic recurrent networks, schematically shown in Figure \ref{fig:spec_clvsa}. The encoder and decoder of the sequence-to-sequence framework take 2-D frames of historical trading data of two consecutive days as input, respectively. The inter-attention module highlights parts of the first one of two consecutive days as the context of the second day. The convolutional LSTM units of the encoder and decoder process 2-D data frames in two steps: i) Convolutional kernels capture local features, ii) Based on the local features, LSTM networks capture temporal features with gated recurrent networks. In each layer of the encoder and decoder, a self-attention module is utilized to highlight parts of the sequence of daily data frames. 

From the above description, we can see that convolutional kernels play a fundamental role in CLVSA. The convolutional kernels operate directly on input data, so the other parts, such as LSTM units and attention layers, work based on local features extracted by convolutional kernels. However, as demonstrated in \cite{wang2018financial}, Cross-Data-Type 1-D Convolution (CDT 1-D Convolution) is applied as convolutional kernels in CLVSA to accommodate the characteristics of historical trading data, which is comprised of five elements: Open, High, Low, Close prices, and Volume. However, there is a prerequisite to use CDT 1-D Convolution, that is, all elements should have strong relevance with each other (e.g. prices and volume under financial markets) so they can share parameters. Our experimental results show that the performance of CLVSA with a direct fusion of TRMI data and historical SPY trading data ($CLVSA^{2}$ in Table \ref{tb:chpt5: result}) degrades by 18.5\%, and 1.01 for average annual return (AAR), Sharpe ratio (SR), respectively, compared to CLVSA with historical SPY trading data only ($CLVSA^{1}$).

\begin{figure*}
\centering
\includegraphics[width=0.9\linewidth]{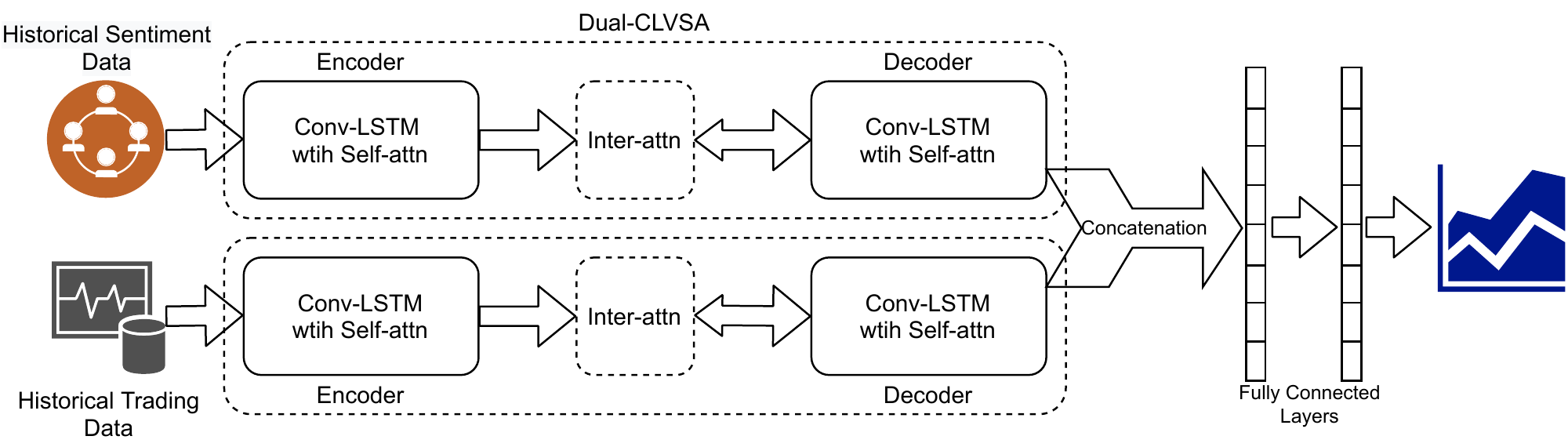}
\caption{The architecture of dual-CLVSA. We add another sequence-to-sequence framework to train historical sentiment data, compared to the original CLVSA. We concatenate the outputs of the two sequence-to-sequence framework before projection layers. We do not apply another set of Kullback-Leibler divergence (KLD) for the channel of sentiment data because of the impulsive characteristic of sentiment data.}
\label{fig:spec_dual_clvsa}
\end{figure*}

To solve this problem, we propose a dual-CLVSA model to fuse TRMI data and historical trading data. The architecture of dual-CLVSA is illustrated in Figure \ref{fig:spec_dual_clvsa}. The basic idea is that, we assign two separate sequence-to-sequence framework to TRMI data and historical trading data, respectively. The two channels are not fused until outputs of decoders from the two channels are concatenated and fed into fully connected layers. On one hand, two separate channels avoid mix-up on convolutions as they have different characteristics; on the other hand, the two channels are combined after individual sequence-to-sequence framework, guaranteeing that the two independent sets of features are processed with the same weight in the fully connected layers. We do not apply another set of Kullback-Leibler divergence (KLD) for the channel of TRMI data because of the sporadic characteristic of sentiment data.

%% file: exp_setting.tex
\section{Experimental setup and Results}\label{Experimental setup}
\subsection{Preprocessing TRMI Data}\label{Data}
We bind TRMI data to raw trading data of the corresponding securities with the same time stamps. That means, we treat TRMI as sentiment ``indicators'', and expect these sentiment indicators to provide the information that is not contained in price movement and trading volume. Specifically,
\begin{enumerate}
    \item The datasets we used in this paper include two parts: (1) historical trading records of a commodity futures WTI Crude Oil (CL), and an exchange-traded fund, SPDR S\&P 500 Trust ETF (SPY). Both of these securities include the following seven attributes: date, time, open price, high price, low price, close price, and trading volume; (2) The corresponding TRMI data, from which we choose the following five common indices as sentiment features: \emph{buzz, sentiment, optimism, fear, and joy}. For the models that contain convolutional kernels, we follow the preprocessing strategy in \cite{wang2018financial}; For others that do not contain convolutional kernels,  we aggregate historical trading records into half-hour data points, and normalize the TRMI data weighted by the Buzz,
    \begin{gather*}
    TRMI_{T}(a) = \frac{\sum_{i \in T}(Buzz_{i}*TRMI_{i}(a))}{\sum_{i \in T}Buzz_{i}},
    \end{gather*}
    where $T$ denotes the desired time interval, which is half hour in our research, $i$ denotes the time stamps within $T$, $a$ denotes the type of TRMI (e.g., joy, fear). After the aggregation, we also bind the two types of data with time stamps.
    \item We guarantee that datasets with and without sentiment measurement are aligned in time for the purpose of meaningful comparisons. After the alignment, the datasets contain both historical trading and sentiment data. We then pick the corresponding fields according to experimental setup. Therefore, the binding procedure is similar to sentiment data ``right joining'' historical trading data with timestamps. We inevitably need to add paddings into sentiment data when TRMI data are missing in some parts of the data. It is a normal operation because of the impulsing characteristic of sentiment data, however, too much padding will harm the prediction performance. More details will be described in Section \ref{chpt6:case2}.
\end{enumerate}

\begin{figure*}
\centering
\includegraphics[width=0.9\linewidth]{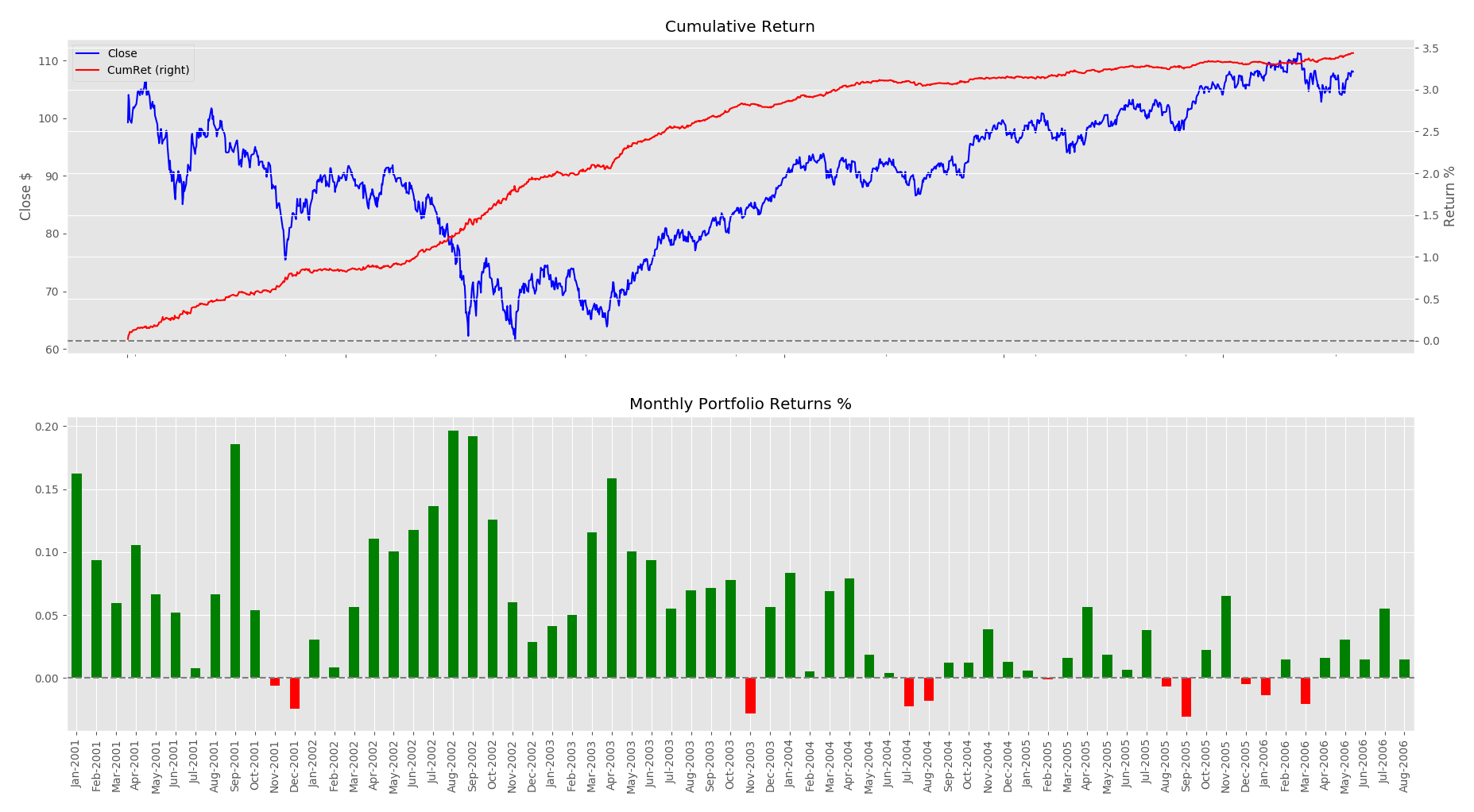}
\caption{The cumulative and monthly return of SPY by dual-CLVSA with historical SPY trading data and TRMI data.}
\label{fig:SPY_dual_CLVSA_trmi}
\end{figure*}

\begin{table}
\centering
    \begin{tabular}{*{6}{c|}}
        \cline{2-6}
         & MAP & AAR & SR & DJA & YJA \\
        \hline
        \multicolumn{1}{|c|}{$LSTM_{s}^{1}$} & $42.0\%$ & $-19.9\%$ & $-1.45$ & $-0.12\%$ & $-21.3\%$  \\
        \multicolumn{1}{|c|}{$LSTM_{s}^{2}$} & $46.0\%$ & $34.2\%$ & $1.75$ & $0.24\%$ & $43.6\%$ \\
        \multicolumn{1}{|c|}{$LSTM_{s}^{3}$} & $45.5\%$ & $32.8\%$ & $1.27$ & $0.24\%$ & $41.9\%$ \\
        \multicolumn{1}{|c|}{$LSTM_{s}^{4}$} & $45.7\%$ & $28.9\%$ & $1.08$ & $0.22\%$ & $37.2\%$ \\
        \hline
        \multicolumn{1}{|c|}{$CLVSA^{1}$} & $46.1\%$ & $48.0\%$ & $2.11$ & $0.24\%$ & $57.9\%$ \\
        \multicolumn{1}{|c|}{$CLVSA^{2}$} & $45.4\%$ & $29.5\%$ & $1.10$ & $0.14\%$ & $35.6\%$ \\
        \hline
        \hline
        \multicolumn{1}{|c|}{dual-CLVSA} & \pmb{$50.7\%$}  & \pmb{$57.3\%$} & \pmb{$3.01$} & \pmb{$0.29\%$} & \pmb{$65.0\%$} \\
        \hline
    \end{tabular}
        \caption{Experimental results of SPY. MAP, AAR, SR, DJA, YJA denote the mean average precision, average annual return, and Sharpe ratio, Daily Jensen Alpha, and Yearly Jensen Alpha respectively.}
\label{tb:chpt5: result}
\end{table}


\subsection{Experimental results of $LSTM_{s}$}
The baseline method at the first step aims to verify the informativeness of sentiment data. We use recurrent neural network with LSTM units ($LSTM_{s}$) to train and test the following four types of datasets, SPY historical trading data, SPY historical trading data with technical indicators, SPY historical trading data with sentiment data, SPY historical trading data with technical indicators and sentiment data. We named the above four experimental sessions as $LSTM^{1}_{s}$ to $LSTM^{4}_{s}$, respectively. Table \ref{tb:chpt5: result} shows their experimental results.

Since $LSTM_{s}$ is designed for temporal feature extraction, which lacks the capability of local feature extraction. Consequently, the experimental results of $LSTM^{1}_{s}$ shows severe losses, -19.9\% average annual return (AAR). $LSTM^{2}_{s}$, however, stays positive and achieves an AAR of 34.2\%. The significant difference between the above two experiments demonstrates that technical indicators provide informative local features to $LSTM_{s}$. 

The experimental results of $LSTM^{3}_{s}$ show a positive AAR of 32.8\% as well. Although it works slightly worse than the experiment with technical indicators, the performance is significantly better than the experiment with historical SPY trading data only ($LSTM^{1}_{s}$). This result verifies that TRMI data is able to provide informative  features as technical indicators.

The experiments of $LSTM^{4}_{s}$ show interesting results. Compared to the aforementioned two experiments, the AAR drops to 28.9\%, indicating that technical indicators and TRMI data can not be fused directly although both of them contain informative features. We also observe similar results in the experiments of CLVSA with SPY historical trading data and TRMI data, which is demonstrated in the next section. 

While the experiments show  the informativeness of TRMI data, the mediocre performance of $LSTM_{s}$ with either TRMI data or the mixed data indicates that $LSTM_{s}$ may not be the optimal framework to take advantages of TRMI data.

\subsection{Experimental results of CLVSA}
The baseline method of the second step aims to reproduce the experimental results as described in \cite{ijcai2019-514} with SPY historical trading data. In \cite{ijcai2019-514}, CLVSA achieves the best performance among the five models for all the datasets of six futures. We thus test the performance of CLVSA on SPY historical trading data, named $CLVSA^{1}$, as shown in Table \ref{tb:chpt5: result}. $CLVSA^{1}$ achieves an AAR of 48.0\% over the same time period, outperforming all the previous experiments with $LSTM_{s}$. This result verifies the superior performances of CLVSA, and thus we choose CLVSA to be the base model for the 3rd-step experiments.

We also investigate the performance of CLVSA with a direct fusion of TRMI data and historical SPY trading data, named $CLVSA^{2}$. We treat TRMI data as alternative ``technical indicators'', in other words, TRMI data is fed into the convolutional kernels of CLVSA along with historical trading data. Similar to $LSTM_{s}^{4}$, $CLVSA^{2}$ underperforms $CLVSA^{2}$ for AAR by 18.5\%, which confirms again that it does not work to fuse historical trading data and TRMI data directly at the input. 

To sum up, the first-step experiments verify that TRMI data is able to provide informative features for price movement prediction, while they also indicate that we can not simply combine TRMI data and historical trading data; the second-step experiments yield similar results to the ones in our previous research, demonstrating again that CLVSA outperforms the singular models such as $LSTM_{s}$. Meanwhile, the results also show that we need a better fusion strategy to take advantage of TRMI data. 

\subsection{Experimental results of dual-CLVSA}
 Figure \ref{fig:SPY_dual_CLVSA_trmi} shows the experimental results of dual-CLVSA with the SPY TRMI data and historical trading data. The cumulative return of SPY remains positive for all the months, and eventually achieves 380\%. The monthly return stay positive for 57 out of 68 months, and not a single month suffers a negative return below -5\%. Daul-CLVSA also exceeds the baseline methods. Compared to $LSTM_{s}$, dual-CLVSA surpasses them for mean average precision (MAP), average annual return (AAR), Sharpe ratio (SR), Daily Jensen Alpha (DJA), and Yearly Jensen Alpha (YJA) by up to 7.3\%, 77.2\%, 4.46, 0.41\%, and 86.3\%, respectively. Compared to CLVSA, dual-CLVSA outperforms it for MAP, AAR, SR, DJA, and YJA by 1.9\%, 24.1\%, 1.30, 0.15\%, and 7.1\%, respectively. Our experimental results verify our hypothesis that with appropriate approach to fusing TRMI data and historical trading data, TRMI data provides extra informative features and thus boost the performance of the predictions and financial returns. 
 
 We explore more about how TRMI data works in dual-CLVSA with the following two cases.

%% file: case_analysis.tex
\section{Case Analysis}\label{Case Analysis}
\subsection{Informativeness of TRMI data in bull and bear markets}
People usually become more emotional when financial markets enter into a bull or bear market. We thus look into two particular time periods, one in a bull market (from May 2003 to March 2004) and the other in a bear market (from March 2002 to July 2002), to investigate the effectiveness of TRMI data for predicting the market movement. 


Our experimental results show that TRMIs are informative for financial markets prediction. Compared to $CLVSA^{1}$ with only SPY historical trading data in the bull market, dual-CLVSA captures 104 more trading opportunities, yields 29\% higher Profitable to Unprofitable Trades ratio, and achieves higher monthly return, Sharpe ratio, and daily Jensen alpha by 46.0\%, 2.84, and 0.19\%, respectively, as shown in Table \ref{tb:chpt6: bull}. In the bear market, dual-CLVSA captures 52 more trading opportunities yields 6\% higher Profitable to Unprofitable Trades, and achieves higher monthly return, Sharpe ratio, and daily Jensen alpha by 12.5\%, 1.62,and 0.11\%, as show in Table \ref{tb:chpt6: bear}. 

\begin{table}[!t]
\centering
    \begin{tabular}{*{6}{c|}}
        \cline{2-6}
         & TC & WT/LT & MR & SR & DJA \\
        \hline
        \multicolumn{1}{|c|}{$CLVSA^{1}$} & $277$ & $154/123$ & $29.3\%$ & $1.58$ & $0.01\%$ \\
        \hline
        \hline
        \multicolumn{1}{|c|}{dual-CLVSA} & \pmb{$381$}  & \pmb{$231/150$} & \pmb{$75.3\%$} & \pmb{$4.42$} & \pmb{$0.20\%$} \\
        \hline
    \end{tabular}

        \caption{Case analysis A: comparison between $CLVSA^{1}$ and dual-CLVSA over the bull market period (from May 2003 to March 2004). TC: trade count, WT/LT: profitable trades/unprofitable trades, MR: monthly return, SR: Sharpe ratio, DJA: daily Jensen alpha. The extra features from TRMI data makes dual-CLVSA capture 104 more trading opportunities and yield $29\%$ higher ratio of Profitable to Unprofitable Trades, outperforming $CLVSA^{1}$ for MR, SR, and DJA by 46.0\%, 2.84, and 0.19\%, respectively.}
\label{tb:chpt6: bull}
\end{table}

\begin{table}[!t]
\centering
    \begin{tabular}{*{6}{c|}}
        \cline{2-6}
         & TC & WT/LT & MR & SR & DJA \\
        \hline
        \multicolumn{1}{|c|}{$CLVSA^{1}$} & $149$ & $83/66$ & $107.3\%$ & $4.78$ & $0.89\%$ \\
        \hline
        \hline
        \multicolumn{1}{|c|}{dual-CLVSA} & \pmb{$201$}  & \pmb{$114/87$} & \pmb{$129.8\%$} & \pmb{$6.40$} & \pmb{$1.00\%$} \\
        \hline
    \end{tabular}
        \caption{Case analysis A: comparison between $CLVSA^{1}$ and dual-CLVSA over the bear market period (from March 2002 to July 2002). TC: trade count, WT/LT: profitable trades/unprofitable trades, MR: monthly return, SR: Sharpe ratio, DJA: daily Jensen alpha. The extra features from TRMI data makes dual-CLVSA capture 52 more trading opportunities and $6\%$ higher ratio of Profitable to Unprofitable Trades, and outperforms $CLVSA^{1}$ for MR, SR, and DJA by 12.5\%, 1.62, and 0.11\%, respectively.}
\label{tb:chpt6: bear}
\end{table}

\subsection{Frequencies of TRMI Data}\label{chpt6:case2}
We also evaluate the performance of dual-CLVSA with crude oil futures (CL) TRMI data and  historical trading data, as shown in Table \ref{tb:chpt6: case2}. Although it achieves average annual return (AAR) of 81.2\% , dual-CLVSA underperforms $CLVSA^{1}$ for mean average precision (MAP), AAR, Sharpe ratio (SR), daily Jensen alpha, and yearly Jensen Alpha by 2.5\%, 31.8\%, 0.98, 0.12\%, and 28.5\% respectively.

\begin{figure}
\centering
  \begin{subfigure}[b]{0.5\textwidth}
    \includegraphics[width=\textwidth]{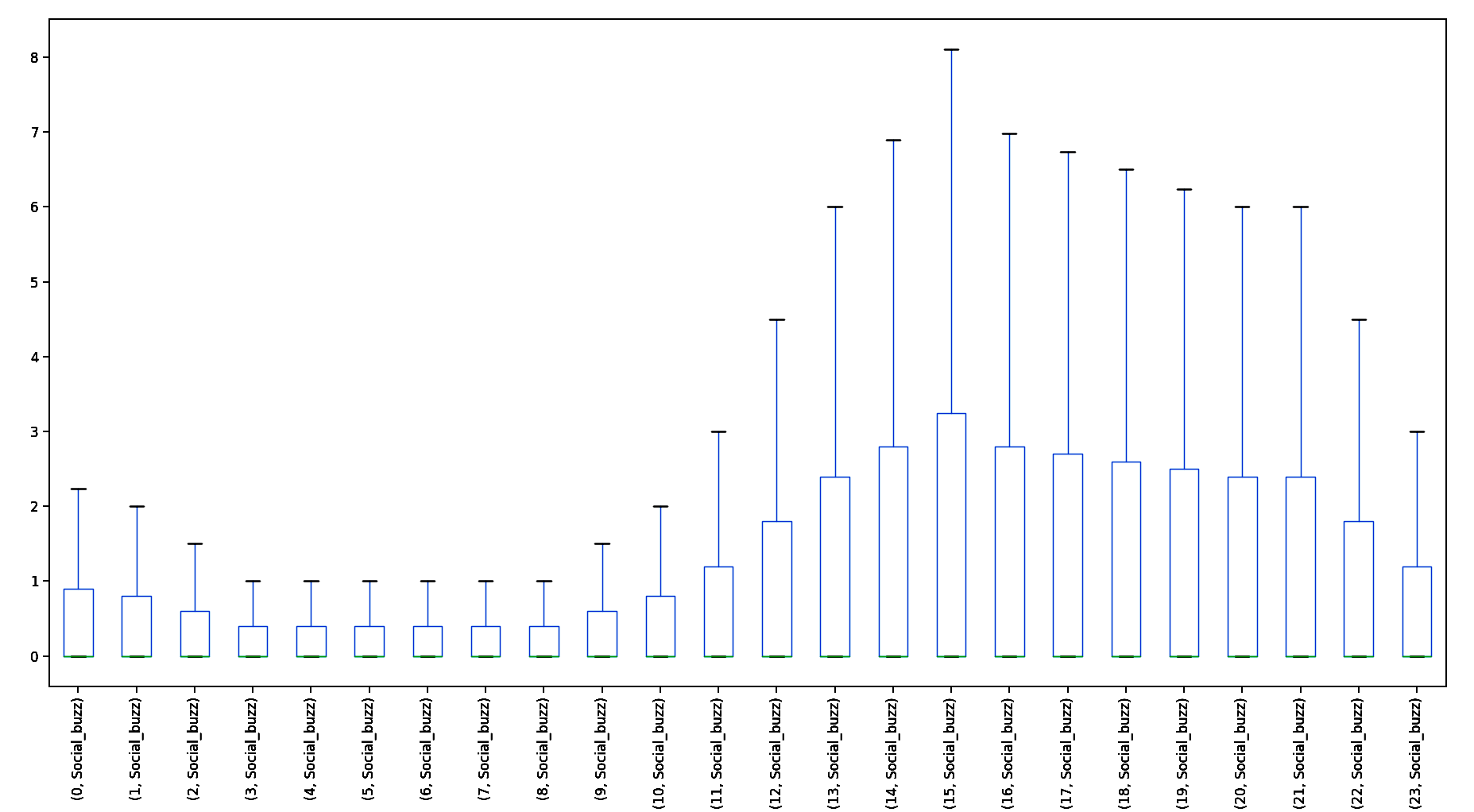}\caption{Hourly CL Social\_buzz boxplots.}
  \end{subfigure}\\
  
  \begin{subfigure}[b]{0.5\textwidth}
    \includegraphics[width=\textwidth]{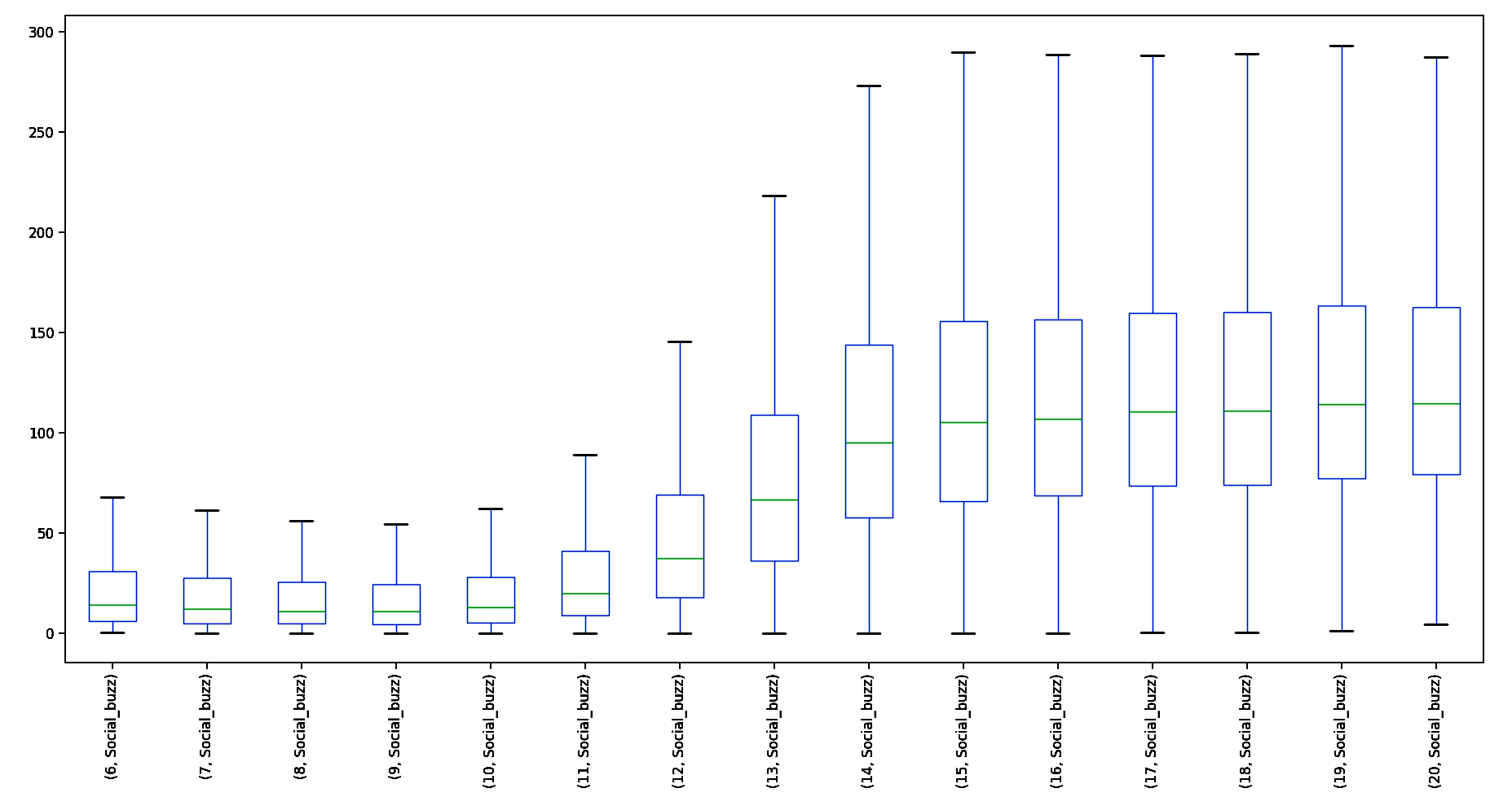}\caption{Hourly SPY Social\_buzz boxplots.}
  \end{subfigure}
  \caption{the Comparisons of Hourly Social\_buzz between CL and SPY.}\label{fig:hourly_trmi}
\end{figure}

To understand why the distinctly different results come out between SPY and CL, we look deeper into TRMI data of SPY and CL. Since Social\_buzz is the weight measurement of TRMI data, it reflects how active social media are at different moments. We plot hourly Social\_buzz boxplots of CL over eight years and SPY over seven years in Figure \ref{fig:hourly_trmi}. While CL's and SPY's Social\_buzz share a similar characteristic that the values in the morning are at a daily low, the Social\_buzz distribution of CL has distinct differences from that of SPY: i) The values of CL Social\_buzz range from 0 to 8, much lower than 0-300 for SPY. ii) The median of CL Social\_buzz in all hours are extremely close to zero, which indicates that almost half of minutely datapoints of CL TRMI data are empty.

\begin{figure*}
\centering
  \begin{subfigure}[b]{1\textwidth}
    \includegraphics[width=\textwidth]{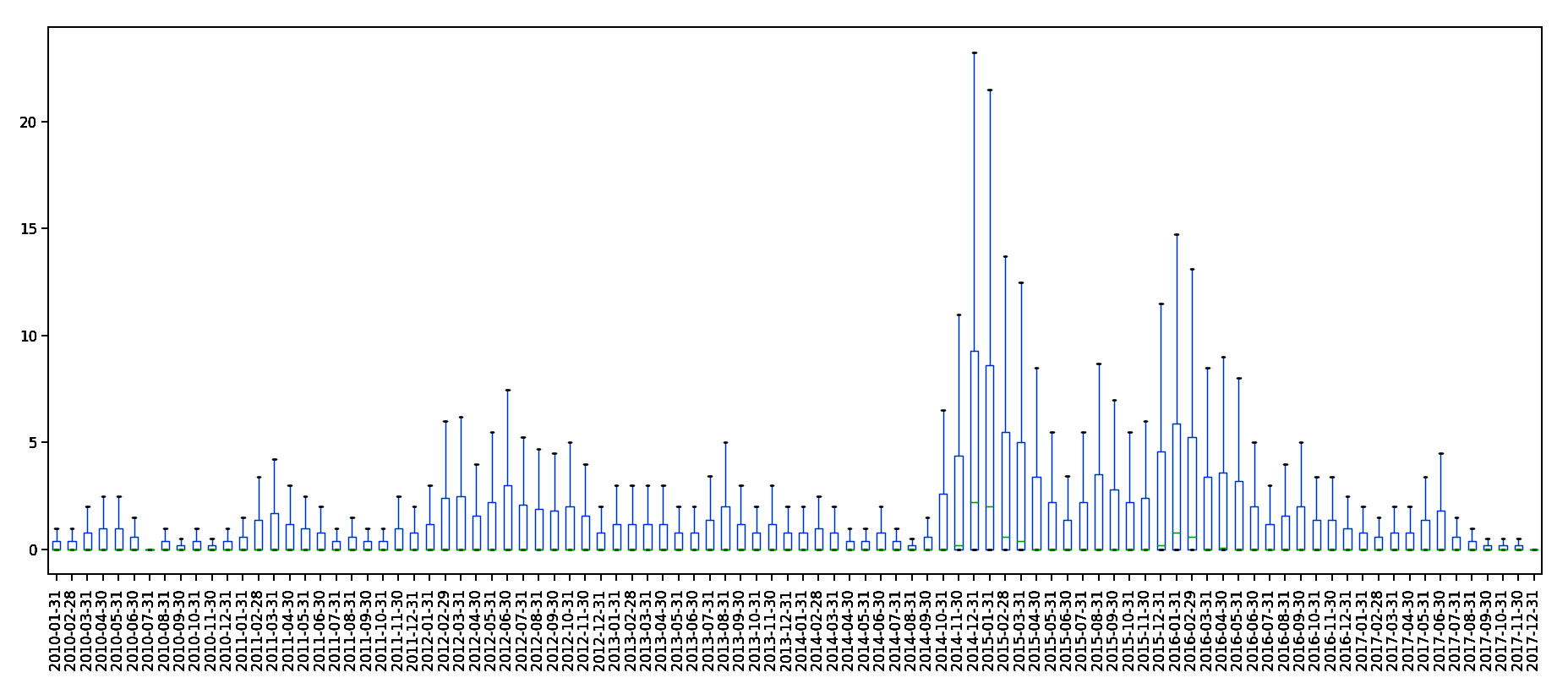}\caption{Calendar-month CL Social\_buzz boxplots from 2010 to 2017.}
  \end{subfigure}\\
  
  \begin{subfigure}[b]{1\textwidth}
    \includegraphics[width=\textwidth]{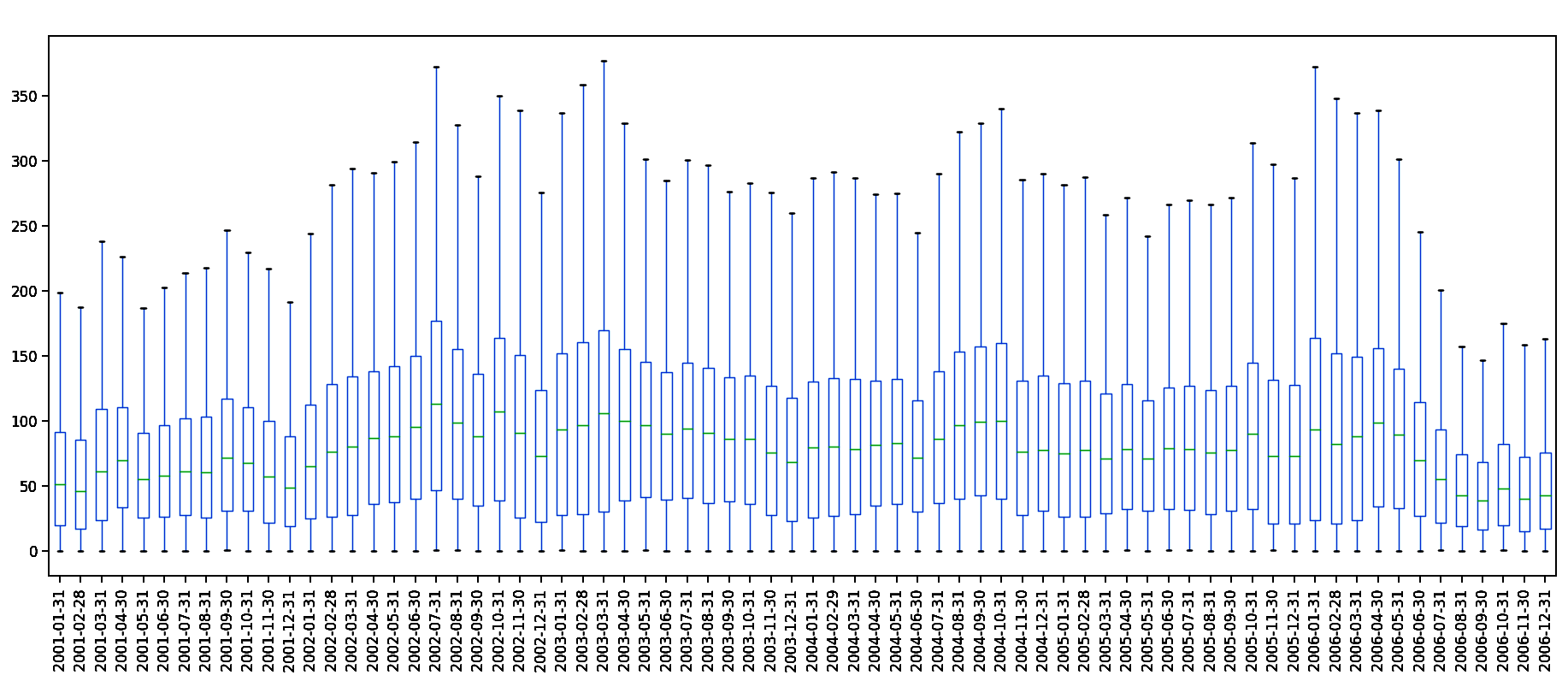}\caption{Calendar-month SPY Social\_buzz boxplots from 2001 to 2006.}
  \end{subfigure}
  \caption{the Comparisons of Monthly Social\_buzz between CL over eight years and SPY over seven years.}\label{fig:month_trmi}
\end{figure*}

We also investigate the calendar-month Social\_buzz distributions of CL and SPY (Figure \ref{fig:month_trmi}). We observe that the calendar-month Social\_buzz distribution of CL display an immerse variability over time. In some months, such as July 2012, social media were completely quiet; while in December 2014 and December 2015, the two months when crude oil prices plummeted, Social\_buzz has a high third quartile and maximum value, even though the first quartile and median values are still very low. From the above analysis, we can see that CL TRMI data is extremely sparse and volatile. Compared to SP500, crude oil futures are much less popular among individual investors, and people discuss about crude oil in social media more sporadically triggered by major events rather than regularly for SP500 which receives a broader interest. The above facts are probably the main reason why the characteristic of CL TRMI data is significantly different from SPY's. The sparsity and volatility of CL TRMI data inevitably result in the poor performance of dual-CLVSA. Specifically, The overly sparse CL TRMI data makes the second channel of sequence-of-sequence framework not able to provide informative features. In other words, the outputs of the second channel may be zero matrices for most of the time, which pollutes the outputs of the first channel after concatenations and thus drag down the overall performance of dual-CLVSA.

\begin{table}[!t]
\centering
    \begin{tabular}{*{6}{c|}}
        \cline{2-6}
         & MAP & AAR & SR & DJA & YJA \\
        \hline
        \multicolumn{1}{|c|}{$CLVSA^{1}$} & \pmb{$49.7\%$} & \pmb{$113.0\%$} & \pmb{$3.99$} & \pmb{$0.55\%$} & \pmb{$107.4\%$} \\
        \hline
        \hline
        \multicolumn{1}{|c|}{dual-CLVSA} & $47.2\%$  & $81.2\%$ & $3.01$ & $0.43\%$ & $78.9\%$ \\
        \hline
    \end{tabular}

        \caption{Case analysis B: comparison between $CLVSA^{1}$ and dual-CLVSA on CL datasets. The overly sparse TRMI data makes dual-CLVSA underperforms $CLVSA^{1}$ for MAP, AAR, SR, DJA, YJA by 2.5\%, 31.8\%, 0.98, 0.12\%, and 28.5\%, respectively.}
\label{tb:chpt6: case2}
\end{table}

%% file: conclusion_futurework.tex
\section{Conclusion}\label{Conclusion}
In this paper, we introduce TRMI data to investigate whether or not the sentiment data provides signals that can help predict financial market movements. Our main contribution is that, based on the state-of-the-art deep learning approach, CLVSA, we design a dual-channel method, named dual-CLVSA, to fuse TRMI data and historical trading data. Our experimental results show that dual-CLVSA outperforms CLVSA by 9.3\% for average annual return and 0.91 for Sharpe ratio on SPDR S\&P 500 ETF Trust. These results indicate that, sentiment data does not only provide informative features to our prediction systems, but also contains extra informative features which prices and volume do not contain. 

%% file: main.bbl
\begin{thebibliography}{10}

\bibitem{kim2003financial}
K.-j. Kim, ``Financial time series forecasting using support vector machines,''
  {\em Neurocomputing}, vol.~55, no.~1, pp.~307--319, 2003.

\bibitem{fernandez2003technical}
F.~Fern{\'a}ndez-Rodr{\'\i}guez, S.~Sosvilla-Rivero, and J.~Andrada-Felix,
  ``Technical analysis in foreign exchange markets: evidence from the ems,''
  {\em Applied Financial Economics}, vol.~13, no.~2, pp.~113--122, 2003.

\bibitem{dixon2016classification}
M.~Dixon, D.~Klabjan, and J.~H. Bang, ``Classification-based financial markets
  prediction using deep neural networks,'' {\em Algorithmic Finance},
  no.~Preprint, pp.~1--11, 2016.

\bibitem{yerkes1908relation}
R.~M. Yerkes, J.~D. Dodson, {\em et~al.}, ``The relation of strength of
  stimulus to rapidity of habit-formation,'' {\em Punishment: Issues and
  experiments}, pp.~27--41, 1908.

\bibitem{hebb1955drives}
D.~O. Hebb, ``Drives and the cns (conceptual nervous system).,'' {\em
  Psychological review}, vol.~62, no.~4, p.~243, 1955.

\bibitem{schwabe2009stress}
L.~Schwabe and O.~T. Wolf, ``Stress prompts habit behavior in humans,'' {\em
  Journal of Neuroscience}, vol.~29, no.~22, pp.~7191--7198, 2009.

\bibitem{peterson2016trading}
R.~L. Peterson, {\em Trading on sentiment: The power of minds over markets}.
\newblock John Wiley \& Sons, 2016.

\bibitem{ijcai2019-514}
J.~Wang, T.~Sun, B.~Liu, Y.~Cao, and H.~Zhu, ``Clvsa: A convolutional lstm
  based variational sequence-to-sequence model with attention for predicting
  trends of financial markets,'' in {\em Proceedings of the Twenty-Eighth
  International Joint Conference on Artificial Intelligence, {IJCAI-19}},
  pp.~3705--3711, International Joint Conferences on Artificial Intelligence
  Organization, 7 2019.

\bibitem{rechenthin2014machine}
M.~D. Rechenthin, ``Machine-learning classification techniques for the analysis
  and prediction of high-frequency stock direction,'' 2014.

\bibitem{ding2015deep}
X.~Ding, Y.~Zhang, T.~Liu, and J.~Duan, ``Deep learning for event-driven stock
  prediction.,'' in {\em Ijcai}, pp.~2327--2333, 2015.

\bibitem{wang2018financial}
J.~Wang, T.~Sun, B.~Liu, Y.~Cao, and D.~Wang, ``Financial markets prediction
  with deep learning,'' in {\em 2018 17th IEEE International Conference on
  Machine Learning and Applications (ICMLA)}, pp.~97--104, IEEE, 2018.

\bibitem{zhang2017stock}
L.~Zhang, C.~Aggarwal, and G.-J. Qi, ``Stock price prediction via discovering
  multi-frequency trading patterns,'' in {\em Proceedings of the 23rd ACM
  SIGKDD International Conference on Knowledge Discovery and Data Mining},
  pp.~2141--2149, ACM, 2017.

\bibitem{bacoyannis2018idiosyncrasies}
V.~Bacoyannis, V.~Glukhov, T.~Jin, J.~Kochems, and D.~R. Song, ``Idiosyncrasies
  and challenges of data driven learning in electronic trading,'' {\em arXiv
  preprint arXiv:1811.09549}, 2018.

\bibitem{xingjian2015convolutional}
S.~Xingjian, Z.~Chen, H.~Wang, D.-Y. Yeung, W.-K. Wong, and W.-c. Woo,
  ``Convolutional lstm network: A machine learning approach for precipitation
  nowcasting,'' in {\em Advances in neural information processing systems},
  pp.~802--810, 2015.

\bibitem{toderici2015variable}
G.~Toderici, S.~M. O'Malley, S.~J. Hwang, D.~Vincent, D.~Minnen, S.~Baluja,
  M.~Covell, and R.~Sukthankar, ``Variable rate image compression with
  recurrent neural networks,'' {\em arXiv preprint arXiv:1511.06085}, 2015.

\bibitem{kaiser2015neural}
{\L}.~Kaiser and I.~Sutskever, ``Neural gpus learn algorithms,'' {\em arXiv
  preprint arXiv:1511.08228}, 2015.

\bibitem{sutskever2014sequence}
I.~Sutskever, O.~Vinyals, and Q.~V. Le, ``Sequence to sequence learning with
  neural networks,'' in {\em Advances in neural information processing
  systems}, pp.~3104--3112, 2014.

\bibitem{bahdanau2014neural}
D.~Bahdanau, K.~Cho, and Y.~Bengio, ``Neural machine translation by jointly
  learning to align and translate,'' {\em arXiv preprint arXiv:1409.0473},
  2014.

\bibitem{cheng2016long}
J.~Cheng, L.~Dong, and M.~Lapata, ``Long short-term memory-networks for machine
  reading,'' {\em arXiv preprint arXiv:1601.06733}, 2016.

\bibitem{kingma2013auto}
D.~P. Kingma and M.~Welling, ``Auto-encoding variational bayes,'' {\em arXiv
  preprint arXiv:1312.6114}, 2013.

\bibitem{rezende2014stochastic}
D.~J. Rezende, S.~Mohamed, and D.~Wierstra, ``Stochastic backpropagation and
  approximate inference in deep generative models,'' {\em arXiv preprint
  arXiv:1401.4082}, 2014.

\bibitem{bayer2014learning}
J.~Bayer and C.~Osendorfer, ``Learning stochastic recurrent networks,'' {\em
  arXiv preprint arXiv:1411.7610}, 2014.

\bibitem{goyal2017z}
A.~G. A.~P. Goyal, A.~Sordoni, M.-A. C{\^o}t{\'e}, N.~R. Ke, and Y.~Bengio,
  ``Z-forcing: Training stochastic recurrent networks,'' in {\em Advances in
  neural information processing systems}, pp.~6713--6723, 2017.

\bibitem{sun2017predicting}
T.~Sun, J.~Wang, P.~Zhang, Y.~Cao, B.~Liu, and D.~Wang, ``Predicting stock
  price returns using microblog sentiment for chinese stock market,'' in {\em
  2017 3rd International Conference on Big Data Computing and Communications
  (BIGCOM)}, pp.~87--96, IEEE, 2017.

\bibitem{xu2018stock}
Y.~Xu and S.~B. Cohen, ``Stock movement prediction from tweets and historical
  prices,'' in {\em Proceedings of the 56th Annual Meeting of the Association
  for Computational Linguistics (Volume 1: Long Papers)}, vol.~1,
  pp.~1970--1979, 2018.

\end{thebibliography}
